\documentclass{article}

 \usepackage[final,nonatbib]{nips_2018}

\usepackage[utf8]{inputenc} 
\usepackage[T1]{fontenc}    
\usepackage{hyperref}       
\usepackage{url}            
\usepackage{booktabs}       
\usepackage{amsfonts}       
\usepackage{nicefrac}       
\usepackage{microtype}      
\usepackage[dvipsnames]{xcolor}
\usepackage{tikz,pgfplots,pgf}
\usepackage{amsmath,graphicx}
\usepackage{mathtools}
\usepackage[utf8]{inputenc} 
\usepackage[T1]{fontenc}    
\usepackage{multirow}
\usepackage{array}
\usepackage{color}
\usepackage{cool}
\usepackage{amssymb}
\usepackage{algorithm}
\usepackage[noend]{algpseudocode}
\usepackage{caption}
\usepackage{wrapfig}

\newcolumntype{P}[1]{>{\centering\let\newline\\\arraybackslash\hspace{0pt}}m{#1}}

\title{Speech Recognition \\ with Quaternion Neural Networks}

\author{
  Titouan Parcollet$^{1,4}$ \quad Mirco Ravanelli$^2$ \quad Mohamed Morchid$^1$ \\ \textbf{Georges Linarès$^1$} \quad \textbf{Renato De Mori$^{1,3}$}\\
  \\
  $^1$LIA, Université d'Avignon, France \\
  $^2$MILA, Université de Montréal, Québec, Canada \\
  $^3$McGill University, Québec, Canada \\
  $^4$ Orkis, Aix-en-provence, France\\
  \texttt{titouan.parcollet@alumni.univ-avignon.fr} \\
  \texttt{mirco.ravanelli@gmail.com}\\
  \texttt{firstname.lastname@univ-avignon.fr}\\
  \texttt{rdemori@cs.mcgill.ca}
}

\begin{document}

\maketitle

%
%
\begin{abstract}

Neural network architectures are at the core of powerful automatic speech recognition systems (ASR). However, while recent researches focus on novel model architectures, the acoustic input features remain almost unchanged. Traditional ASR systems rely on multidimensional acoustic features such as the Mel filter bank energies alongside with the first, and second order derivatives to characterize time-frames that compose the signal sequence. Considering that these components describe three different views of the same element, neural networks have to learn both the internal relations that exist within these features, and external or global dependencies that exist between the time-frames. Quaternion-valued neural networks (QNN), recently received an important interest from researchers to process and learn such relations in multidimensional spaces. Indeed, quaternion numbers and QNNs have shown their efficiency to process multidimensional inputs as entities, to encode internal dependencies, and to solve many tasks with up to four times less learning parameters than real-valued models. We propose to investigate modern quaternion-valued models such as convolutional and recurrent quaternion neural networks in the context of speech recognition with the TIMIT dataset. The experiments show that QNNs always outperform real-valued equivalent models with way less free parameters, leading to a more efficient, compact, and expressive representation of the relevant information.

\end{abstract}

%
%
\section{Introduction}

During the last decade, deep neural networks (DNN) have encountered a wide success in automatic speech recognition. Many architectures such as recurrent (RNN) \cite{sak2014long, hinton2012deep,abdel2012applying,mirco2017timit,greff2017lstm}, time-delay (TDNN) \cite{waibel1990phoneme,peddinti2015time}, or convolutional neural networks (CNN) \cite{zhang2017towards} have been proposed and achieved better performances than traditional hidden Markov models (HMM) combined with gaussian mixtures models (GMM) in different speech recognition tasks. However, despite such evolution of models and paradigms, the acoustic feature representation remains almost the same. The acoustic signal is commonly split into time-frames, for which Mel-filter banks energies, or Mel frequency scaled cepstral coefficients (MFCC) \cite{davis1990comparison} are extracted, alongside with the first and second order derivatives. In fact, time-frames are characterized by $3$-dimensional features that are related by representing three different views of the same basic element. Consequently an efficient neural networks-based model has to learn both external dependencies between time-frames, and internal relations within the features. Traditional real-valued architectures deal with both dependencies at the same level, due to the lack of a dedicated mechanism to learn the internal and external relations separately. 

Quaternions are hypercomplex numbers that contain a real and three separate imaginary components, fitting perfectly to three and four dimensional feature vectors, such as for image processing and robot kinematics \cite{sangwine1996fourier,pei1999color,aspragathos1998comparative}. The idea of bundling groups of numbers into separate entities is also exploited by the recent capsule network ~\cite{hinton2017capsule}. Contrary to traditional homogeneous representations, capsule and quaternion neural networks bundle sets of features together. Thereby, quaternion numbers allow neural models to code latent inter-dependencies between groups of input features during the learning process with up to four times less parameters than real-valued neural networks, by taking advantage of the {\em Hamilton product} as the
equivalent of the dot product between quaternions. Early applications of quaternion-valued backpropagation algorithms \cite{arena1994neural,arena1997multilayer} have efficiently solved quaternion functions approximation tasks. More recently, neural networks of complex and hypercomplex numbers have received an increasing attention \cite{hirose2012generalization,tygert2016mathematical,danihelka2016associative,wisdom2016full}, and some efforts have shown promising results in different applications. In particular, a deep quaternion network \cite{parcollet2016quaternion, parcollet2017deep, parcollet2017quaternion}, and a deep quaternion convolutional network \cite{chase2017quat,parcollet2018quaternion} have been successfully employed for challenging tasks such as images and language processing. 

\textbf{Contributions:} This paper proposes to evaluate previously investigated quaternion-valued models in two different realistic conditions of speech recognition, to see whether the quaternion encoding of the signal, alongside with the quaternion algebra and the important parameter reduction help to better capture the acoustic signal nature, leading to a more expressive representation of the information. Based on the TIMIT \cite{garofolo1993darpa} phoneme recognition task, a quaternion convolutional neural network (QCNN) is compared to a real-valued CNN in a end-to-end framework, and a quaternion recurrent neural network (QRNN) is compared to an RNN within a more traditional HMM-based system. In the end-to-end approach, the experiments show that the QCNN outperforms the CNN with a phoneme error rate (PER) of $19.5\%$ against the $20.6\%$ achieved for CNNs. Moreover, the QRNN outperforms the RNN with a PER of $18.5\%$ against $19.0\%$ for the RNN. Furthermore, such results are observed with a maximum reduction factor of the number of neural network parameters of $3.96$ times.

%
%
\section{Motivations}

A major challenge of current machine learning models is to obtain efficient representations of relevant information for solving a specific task. Consequently, a good model has to efficiently code both the relations that occur at the feature level, such as between the Mel filter energies, the first, and second order derivatives values of a single time-frame, and at a global level, such as a phonemes or words described by a group of time-frames. Moreover, to avoid overfitting, better generalize, and to be more efficient, such models also have to be as small as possible. Nonetheless, real-valued neural networks usually require a huge set of parameters to well-perform on speech recognition tasks, and hardly code internal dependencies within the features, since they are considered at the same level as global dependencies during the learning. In the following, we detail the motivations to employ quaternion-valued neural networks instead of real-valued ones to code inter and intra features dependencies with less parameters.
 
\begin{figure}[!t]
    \centering
    \includegraphics[scale=0.28]{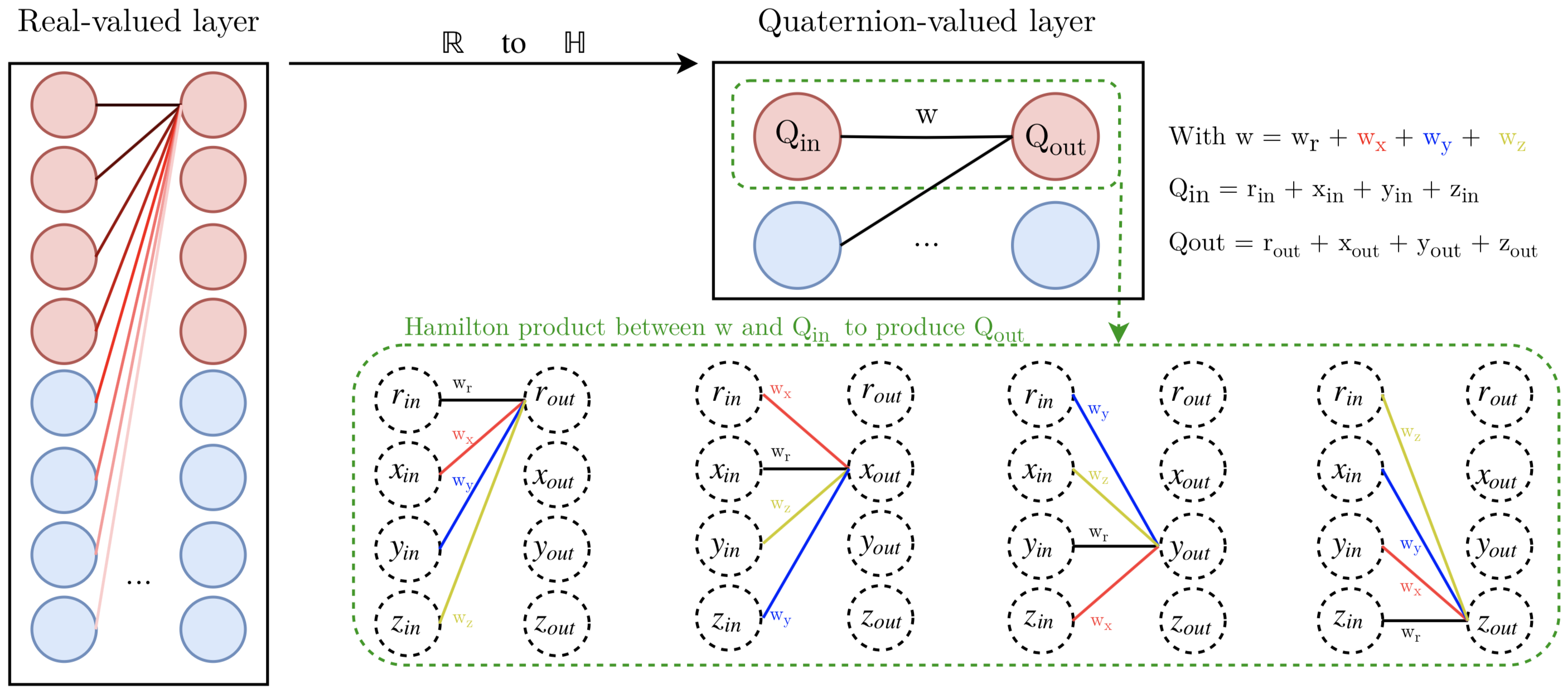}
    \caption{Illustration of the input features ($Q_{in}$) latent relations learning ability of a quaternion-valued layer (right) due to the quaternion weight sharing of the {\em Hamilton product} (Eq. \ref{eq:hamilton}), compared to a standard real-valued layer (left).}
    \label{fig:proof}
\end{figure}

First, a better representation of multidimensional data has to be explored to naturally capture internal relations within the input features. For example, an efficient way to represent the information composing an acoustic signal sequence is to consider each time-frame as being a whole entity of three strongly related elements, instead of a group of unidimensional elements that \textit{could} be related to each others, as in traditional real-valued neural networks. Indeed, with a real-valued NN, the latent relations between the Mel filter banks energies, and the first and second order derivatives of a given time-frame are hardly coded in the latent space since the weight has to find out these relations among all the time-frames composing the sequence. Quaternions are fourth dimensional entities and allow one to build and process elements made of up to four elements, mitigating the above described problem. Indeed, the quaternion algebra and more precisely the {\em Hamilton product} allows quaternion neural network to capture these internal latent relations within the features of a quaternion. It has been shown that QNNs are able to restore the spatial relations within $3$D coordinates \cite{matsui2004quaternion}, and within color pixels \cite{isokawa2003quaternion}, while real-valued NNs failed. In fact, the quaternion-weight components are shared through multiple quaternion input parts during the {\em Hamilton product }, creating relations within the elements. Indeed, Figure \ref{fig:proof} shows that the multiple weights required to code latent relations within a feature are considered at the same level as for learning global relations between different features, while the quaternion weight $w$ codes these internal relations within a unique quaternion $Q_{out}$ during the {\em Hamilton product} (right).

Second, quaternion neural networks make it possible to deal with the same signal dimension than real-valued NN, but with four times less neural parameters. Indeed, a $4$-number quaternion weight linking two 4-number quaternion units only has $4$ degrees of freedom, whereas a standard neural net parametrization have $4 \times 4=16$, i.e., a 4-fold saving in memory. Therefore, the natural multidimensional representation of quaternions alongside with their ability to drastically reduce the number of parameters indicate that hyper-complex numbers are a better fit than real numbers to create more efficient models in multidimensional spaces such as speech recognition. 

Indeed, modern automatic speech recognition systems usually employ input sequences composed of multidimensional acoustic features, such as log Mel features, that are often enriched with their first, second and third time derivatives \cite{davis1990comparison, furui1986speaker}, to integrate contextual information. In standard NNs, static features are simply concatenated with their derivatives to form a large input vector, without effectively considering that signal derivatives represent different views of the same input. Nonetheless, it is crucial to consider that these three descriptors represent a special state of a time-frame, and are thus correlated. Following the above motivations and the results observed on previous works about quaternion neural networks, we hypothesize that for acoustic data, quaternion NNs naturally provide a more suitable representation of the input sequence, since these multiple views can be directly embedded in the multiple dimensions space of the quaternion, leading to smaller and more accurate models. 

%
%
\section{Quaternion Neural Networks}

Real-valued neural networks architectures are extended to the quaternion domain to benefit from its capacities. Therefore, this section proposes to introduce the quaternion algebra (Section \ref{subsec:qalgebra}), the quaternion internal representation (section \ref{subsec:internal}), a quaternion convolutional neural networks (QCNN, Section \ref{subsec:qcnn}) and a quaternion recurrent neural network (QRNN, Section \ref{subsec:qrnn}).

%
%
\subsection{Quaternion Algebra}
\label{subsec:qalgebra}
The quaternion algebra $\mathbb{H}$ defines operations between quaternion numbers. A quaternion Q is an extension of a complex number defined in a four dimensional space as:
\begin{align}
Q = r1 + x\textbf{i} + y\textbf{j} + z\textbf{k},
\end{align}
where $r$, $x$, $y$, and $z$ are real numbers, and $1$, \textbf{i}, \textbf{j}, and \textbf{k} are the quaternion unit basis. In a quaternion, $r$ is the real part, while $x\textbf{i}+y\textbf{j}+z\textbf{k}$ with $\textbf{i}^2=\textbf{j}^2=\textbf{k}^2=\textbf{i}\textbf{j}\textbf{k}=-1$ is the imaginary part, or the vector part. 
Such a definition can be used to describe spatial rotations. The information embedded in the quaterion $Q$ can be summarized into the following matrix of real numbers, that turns out to be more suitable for computations:
\begin{align}
Q_{mat} = 
\begin{bmatrix}
   r & -x & -y & -z \\
   x & r & -z & y \\
   y & z & r & -x \\
   z & -y & x & r 
\end{bmatrix}.
\end{align}
The conjugate $Q^*$ of $Q$ is defined as:
\begin{align}
\label{eq:conjugate}
Q^*=r1-x\textbf{i}-y\textbf{j}-z\textbf{k}.
\end{align}
Then, a normalized or unit quaternion $Q^\triangleleft$ is expressed as:
\begin{align}
\label{eq:normalize}
Q^\triangleleft=\frac{Q}{\sqrt{r^2+x^2+y^2+z^2}}.
\end{align}
Finally, the Hamilton product $\otimes$ between two quaternions $Q_1$ and $Q_2$ is computed as follows: 
\begin{align}
Q_1 \otimes Q_2=&(r_1r_2-x_1x_2-y_1y_2-z_1z_2)+\nonumber \\
			&(r_1x_2+x_1r_2+y_1z_2-z_1y_2) \boldsymbol i+\nonumber \\
            &(r_1y_2-x_1z_2+y_1r_2+z_1x_2) \boldsymbol j+\nonumber \\
            &(r_1z_2+x_1y_2-y_1x_2+z_1r_2) \boldsymbol k.
\label{eq:hamilton}
\end{align}
The Hamilton product is used in QRNNs to perform transformations of vectors representing quaternions, as well as scaling and interpolation between two rotations following a geodesic over a sphere in the $\mathbb{R}^3$ space as shown in~\cite{minemoto2017feed}.

%
%
\subsection{Quaternion internal representation}
\label{subsec:internal}

In a quaternion layer, all parameters are quaternions, including inputs, outputs, weights, and biases. The quaternion algebra is ensured by manipulating matrices of real numbers \cite{parcollet2018quaternion}. Consequently, for each input vector of size $N$, output vector of size $M$, dimensions are split into four parts: the first one equals to $r$, the second is $x\textbf{i}$, the third one equals to $y\textbf{j}$, and the last one to $z\textbf{k}$ to compose a quaternion $Q = r1+x\textbf{i}+y\textbf{j}+z\textbf{k}$. The inference process is based in the real-valued space on the dot product between input features and weights matrices. In any quaternion-valued NN, this operation is replaced with the {\em Hamilton product} (eq. \ref{eq:hamilton}) with quaternion-valued matrices (i.e. each entry in the weight matrix is a quaternion).

%
%
\subsection{Quaternion convolutional neural networks}
\label{subsec:qcnn}
Convolutional neural networks (CNN) \cite{lecun1999object} have been proposed to capture the high-level relations that occur between neighbours features, such as shape and edges on an image. However, internal dependencies within the features are considered at the same level than these high-level relations by real-valued CNNs, and it is thus not guaranteed that they are well-captured. In this extend, a quaternion convolutional neural network (QCNN) have been proposed by \cite{chase2017quat, parcollet2018quaternion}\footnote{\url{https://github.com/Orkis-Research/Pytorch-Quaternion-Neural-Networks}}. Let $\gamma_{ab}^l$ and $S_{ab}^l$ , be the quaternion output and the pre-activation quaternion output at layer $l$ and at the indexes $(a,b)$ of the new feature map, and $w$ the quaternion-valued weight filter map of size $K \times K$. A formal definition of the convolution process is: 

\begin{align}
\gamma_{ab}^l &=\alpha(S_{ab}^l),
\end{align}
with
\begin{align}
\label{eq:forward}
S_{ab}^l&=\sum\limits_{c=0}^{K-1}\sum\limits_{d=0}^{K-1}w^l \otimes \gamma_{(a+c)(b+d)}^{l-1},
\end{align}
where $\alpha$ is a {\em quaternion split activation} function \cite{xu2017learning} defined as:
\begin{equation}
\label{eq:split}
\alpha(Q)=f(r)+f(x)\textbf{i}+f(y)\textbf{j}+f(z)\textbf{k},
\end{equation}
with $f$ corresponding to any standard activation function. The output layer of a quaternion neural network is commonly either quaternion-valued such as for quaternion approximation \cite{arena1997multilayer}, or real-valued to obtains a posterior distribution based on a softmax function following the split approach of Eq. \ref{eq:split}. Indeed, target classes are often expressed as real numbers. Finally, the full derivation of the backpropagation algorithm for quaternion valued neural networks can be found in \cite{nitta1995quaternary}.

%
%
\subsection{Quaternion recurrent neural networks}
\label{subsec:qrnn}
Despite the fact that CNNs are efficient to detect and learn patterns in an input volume, recurrent neural networks (RNN) are more adapted to represent sequential data. Indeed, recurrent neural networks obtained state-of-the-art results on many tasks related to speech recognition \cite{ravanelli2018light,graves2013speech}. Therefore, a quaternary version of the RNN called QRNN have been proposed by \cite{parcollet2018QRNN} \footnote{\url{https://github.com/Orkis-Research/Pytorch-Quaternion-Neural-Networks}}. Let us define a QRNN with an hidden state composed of $H$ neurons. Then, let $w_{hh}$, $w_{h\gamma}$, and $w_{\gamma h}$ be the hidden to hidden, input to hidden, and hidden to output weight matrices respectively, and $b_{n}^l$ be the bias at neuron $n$ and layer $l$.  Therefore, and with the same parameters as for the QCNN, the hidden state $h_n^{t,l}$ of the neuron $n$ at timestep $t$ and layer $l$ can be computed as:

\begin{align}
h_{n}^{t,l} =\alpha(\sum\limits_{m=0}^{H} w_{nm,hh}^{t,l} \otimes h_{m}^{t-1,l} + \sum\limits_{m=0}^{N_{l-1}} w_{nm,h\gamma}^{t,l} \otimes \gamma_{m}^{t,l-1} + b_{n}^l),
\end{align}
with $\alpha$ any \textit{split activation} function. Finally, the output of the neuron $n$ is computed following:
\begin{align}
\gamma_{n}^{t,l} =\beta(\sum\limits_{m=0}^{N_{l-1}} w_{nm,\gamma h}^{t,l} \otimes h_{n}^{t,l-1} + b_{n}^l),
\end{align}
with $\beta$ any \textit{split activation} function. The full derivation of the backpropagation trough time of the QRNN can be found in \cite{parcollet2018QRNN}.

%
%
\section{Experiments on TIMIT}

Quaternion-valued models are compared to their real-valued equivalents on two different benchmarks with the TIMIT phoneme recognition task \cite{garofolo1993darpa}. First, an end-to-end approach is investigated based on QCNNs compared to CNNs in Section \ref{subsec:endtoend}. Then, a more traditional and powerful method based on QRNNs compared to RNNs alongside with HMM decoding is explored in Section \ref{subsec:hmm}. During both experiments, the training process is performed on the standard $3,696$ sentences uttered by $462$ speakers, while testing is conducted on $192$ sentences uttered by $24$ speakers of the TIMIT dataset. A validation set composed of $400$ sentences uttered by $50$ speakers is used for hyper-parameter tuning. All the results are from an average of three runs (3-folds) to alleviate any variation due to the random initialization.

%
%
\subsection{Acoustic quaternions}

End-to-end and HMM based experiments share the same quaternion input vector extracted from the acoustic signal. The raw audio is first transformed into $40$-dimensional log Mel-filter-bank coefficients using the \textit{pytorch-kaldi}\footnote{\url{https://github.com/mravanelli/pytorch-kaldi}} toolkit and the Kaldi s5 recipes \cite{Povey2011ASRU}. Then, the first, second, and third order derivatives are extracted. Consequently, an acoustic quaternion $Q(f,t)$ associated with a frequency $f$ and a time-frame $t$ is formed as:
\begin{align}
Q(f,t) = e(f,t) + \frac{\partial e(f,t)}{\partial t}\textbf{i} + \frac{\partial^2 e(f,t)}{\partial^2 t} \textbf{j} + \frac{\partial^3 e(f,t)}{\partial^3 t} \textbf{k}.
\end{align}
$Q(f,t)$ represents multiple views of a frequency $f$ at time frame $t$, consisting of the energy $e(f,t)$ in the filter band at frequency $f$, its first time derivative describing a slope view, its second time derivative describing a concavity view, and the third derivative describing the rate of change of the second derivative. Quaternions are used to learn the spatial relations that exist between the different views that characterize a same frequency. Thus, the quaternion input vector length is $160/4 = 40$.

%
%
\subsection{Toward end-to-end phonemes recognition}
\label{subsec:endtoend}

End-to-end systems are at the heart of modern researches in the speech recognition domain \cite{zhang2017towards}. The task is particularly difficult due to the differences that exists between the raw or pre-processed acoustic signal used as input features, and the word or phonemes expected at the output. Indeed, both features are not defined at the same time-scale, and an automatic alignment method has to be defined. This section proposes to evaluate the QCNN compared to traditional CNN in an end-to-end model based on the connectionist temporal classification (CTC) method, to see whether the quaternion encoding of the signal, alongside with the quaternion algebra, help to better capture the acoustic signal nature and therefore better generalize.

%
%
\subsubsection{Connectionist Temporal Classification}

In the acoustic modeling part of ASR systems, the task of sequence-to-sequence mapping from an input acoustic signal $X=[x_1,...,x_n]$ to a sequence of symbols $T=[t_1,...,t_m]$ is complex due to:
\begin{itemize}
\item $X$ and $T$ could be in arbitrary length.
\item The alignment between $X$ and $T$ is unknown in most cases.
\end{itemize}
Specially, $T$ is usually shorter than $X$ in terms of phoneme symbols. To alleviate these problems, connectionist temporal classification (CTC) has been proposed \cite{graves2006connectionist}. First, a softmax is applied at each timestep, or frame, providing a probability of emitting each symbol $X$ at that timestep. This probability results in a symbol sequences representation $P(O|X)$, with $O = [o_1,...,o_n]$ in the latent space $O$. A blank symbol $'-'$ is introduced as an extra label to allow the classifier to deal with the unknown alignment. Then, $O$ is transformed to the final output sequence with a many-to-one function $g(O)$ defined as follows:
\begin{align}
\left.
    \begin{array}{ll}
        g(z_1,z_2,-,z_3,-) \\
        g(z_1,z_2,z_3,z_3,-) \\
        g(z_1,-,z_2,z_3,z_3) \\
    \end{array}
\right \}=(z_1,z_2,z_3). 
\end{align}
Consequently, the output sequence is a summation over the probability of all possible alignments between $X$ and $T$ after applying the function $g(O)$. Accordingly to \cite{graves2006connectionist} the parameters of the models are learned based on the cross entropy loss function: 
\begin{align}
 \sum\nolimits_{X, T \in train}-\log(P(O|X)). 
\end{align}
During the inference, a best path decoding algorithm is performed. Therefore, the latent sequence with the highest probability is obtained by performing argmax of the softmax output at each timestep. The final sequence is obtained by applying the function $g(.)$ to the latent sequence.

%
%
\subsubsection{Model Architectures}

A first $2$D convolutional layer is followed by a maxpooling layer along the frequency axis to reduce the internal dimension. Then, $n=10$ $2$D convolutional layers are included, together with three dense layers of sizes $1024$ and $256$ respectively for real- and quaternion-valued models. Indeed, the output of a dense quaternion-valued layer has $256 \times 4 = 1024$ nodes and is four times larger than the number of units. The filter size is rectangular $(3,5)$, and a padding is applied to keep the sequence and signal sizes unaltered. The number of feature maps varies from $32$ to $256$ for the real-valued models and from $8$ to $64$ for quaternion-valued models. Indeed, the number of output feature maps is four times larger in the QCNN due to the quaternion convolution, meaning $32$ quaternion-valued feature maps (FM) correspond to $128$ real-valued ones. Therefore, for a fair comparison, the number of feature maps is represented in the real-valued space (e.g., a number of real-valued FM of $256$ corresponds to $256/4=64$ quaternion-valued neurons). The PReLU activation function is employed for both models~\cite{he2015delving}. A dropout of $0.2$ and a $L_2$ regularization of $1e^{-5}$ are used across all the layers, except the input and output ones. CNNs and QCNNs are trained with the RMSPROP learning rate optimizer and vanilla hyperparameters~\cite{kingma2014adam} during $100$ epochs. The learning rate starts at $8\cdot10^{-4}$, and is decayed by a factor of $0.5$ every time the results observed on the validation set do not improve. Quaternion parameters including weights and biases are initialized following the adapted quaternion initialization scheme provided in \cite{parcollet2018QRNN}. Finally, the standard CTC loss function defined in~\cite{graves2006connectionist} and implemented in \cite{chollet2015keras} is applied.

%
%
\subsubsection{Results and discussions}

End-to-end results of QCNN and CNN are reported in Table \ref{table:timit1}. In agreement with our hypothesis, one may notice an important difference in the amount of learning parameters between real and quaternion valued CNNs. An explanation comes from the quaternion algebra. A dense layer with $1,024$ input values and $1,024$ hidden units contains $1,024^2\approx1$M parameters, while the quaternion equivalent needs $256^2\times4\approx0.26$M parameters to deal with the same signal dimension. Such reduction in the number of parameters have multiple positive impact on the model. First, a smaller memory footprint for embedded and limited devices. Second, and as demonstrated in Table \ref{table:timit1}, a better generalization ability leading to better performances. Indeed, the best PER observed in realistic conditions (w.r.t to the development PER) is $19.5\%$ for the QCNN compared to $20.6\%$ for the CNN, giving an absolute improvement of $0.5\%$ with QCNN. Such results are obtained with $32.1$M parameters for CNN, and only $8.1M$ for QCNN, representing a reduction factor of $3.96$x of the number of parameters. 

\begin{table}[!h]
    \caption{Phoneme error rate (PER\%) of CNN and QCNN  models on the development and test sets of the TIMIT dataset. “Params" stands for the total number of trainable parameters, and "FM" for the feature maps size.}
        \centering
        \scalebox{1}{
        \begin{tabular}{ P{1.3cm}P{1.3cm}  P{1cm} P{1cm}  P{1.3cm}}
           \hline\hline
            \textbf{Models}& \textbf{FM}&\textbf{Dev.}& \textbf{Test}& \textbf{Params}\\
            \hline
        
            \multirow{4}*{CNN}  & 32  & 22.0  & 23.1 & 3.4M \\
                                & 64  & 19.6  & 20.7 & 5.4M \\
           	                    & 128 & 19.6  & 20.8 & 11.5M \\
                                & \textbf{256} & \textbf{19.0}  & \textbf{20.6} & \textbf{32.1M} \\
           	\hline
           	\multirow{4}*{QCNN} & 32  & 22.3 & 23.3  &  0.9M \\
                                & 64  & 19.9 & 20.5  &  1.4M \\
           	                    & 128 & 18.9 & 19.9  &  2.9M \\
                                & \textbf{256} & \textbf{18.2} & \textbf{19.5}  &  \textbf{8.1M} \\
           	\hline
        \end{tabular}
        \label{table:timit1}
        }
\end{table}

It is worth noticing that with much fewer learning parameters for a given architecture, the QCNN always performs better than the real-valued one. Consequently, the quaternion-valued convolutional approach offers an alternative to traditional real-valued end-to-end models, that is more efficient, and more accurate. However, due to an higher number of computations involved during the { \em Hamilton product } and to the lack of proper engineered implementations, the QCNN is one time slower than the CNN to train. Nonetheless, such behavior can be alleviated with a dedicated implementation in CUDA of the { \em Hamilton product }. Indeed, this operation is a matrix product and can thus benefit from the parallel computation of GPUs. In fact, a proper implementation of the {\em Hamilton product} will leads to a higher and more efficient usage of GPUs.

%
%
\subsection{HMM-based phonemes recognition}
\label{subsec:hmm}

A conventional ASR pipeline based on a HMM decoding process alongside with recurrent neural networks is also investigated to reach state-of-the-art results on the TIMIT task. While input features remain the same as for end-to-end experiments, RNNs and QRNNs are trained to predict the HMM states that are then decoded in the standard Kaldi recipes \cite{Povey2011ASRU}. As hypothetized for the end-to-end solution, QRNN models are expected to better generalize than RNNs due to their specific algebra.

%
%
\subsubsection{Model Architectures}

RNN and QRNN models are compared on a fixed number of layers $M=4$ and by varying the number of neurons $N$ from $256$ to $2,048$, and $64$ to $512$ for the RNN and QRNN respectively. Indeed, as demonstrated on the previous experiments the number of hidden neurons in the quaternion and real spaces do not handle the same amount of real-number values. Tanh activations are used across all the layers except for the output layer that is based on a softmax function. Models are optimized with RMSPROP \cite{kingma2014adam} with vanilla hyper-parameters and an initial learning rate of $8\cdot10^{-4}$. The learning rate is progressively annealed using an halving factor of $0.5$ that is applied when no performance improvement on the validation set is observed. The models are trained during $25$ epochs. A dropout rate of $0.2$ is applied over all the hidden layers \cite{srivastava2014dropout} except the output. The negative log-likelihood loss function is used as an objective function. As for QCNNs, quaternion parameters are initialized based on \cite{parcollet2018QRNN}. Finally, decoding is based on Kaldi \cite{Povey2011ASRU} and weighted finite state transducers (WFST) \cite{MOHRI2002mohri} that integrate acoustic, lexicon and language model probabilities into a single HMM-based search graph.

%
%
\subsubsection{Results and discussions}

The results of both QRNNs and RNNs alongside with an HMM decoding phase are presented in Table \ref{table:timit2}. A best testing PER of $18.5\%$ is reported for QRNN compared to $19.0\%$ for RNN, with respect to the best development PER. Such results are obtained with $3.8$M, and $9.4$M parameters for the QRNN and RNN respectively, with an equal hidden dimension of $1,024$, leading to a reduction of the number of parameters by a factor of $2.47$ times. As for previous experiments the QRNN always outperform equivalents architectures in term of PER, with significantly less learning parameters. It is also important to notice that both models tend to overfit with larger architectures. However, such phenomenom is lowered by the small number of free parameters of the QRNN. Indeed, a QRNN whose hidden dimension is $2,048$ only has $11.2$M parameters compared to $33.4$M for an equivalently sized RNN, leading to less degrees of freedom, and therefore less overfitting.  

\begin{table}[!h]

    \caption{Phoneme error rate (PER\%) of both QRNN, RNN  models on the development and test sets of the TIMIT dataset. “Params" stands for the total number of trainable parameters, and "Hidden dim" represents the dimension of the hidden layers. As an example, a QRNN layer of "Hidden dim." $=512$ is equivalent to $128$ hidden quaternion neurons.}
    
        \centering
        \scalebox{1}{
        \begin{tabular}{ P{1.3cm}P{1.5cm}  P{1cm} P{1cm}  P{1.3cm}}
           \hline\hline
            \textbf{Models}& \textbf{Hidden dim.}&\textbf{Dev.}& \textbf{Test}& \textbf{Params}\\
            \hline
        
            \multirow{4}*{RNN}  & 256  & 22.4  & 23.4 & 1M \\
                                & 512  & 19.6  & 20.4 & 2.8M \\
           	                    & \textbf{1,024} & \textbf{17.9}  & \textbf{19.0} & \textbf{9.4M} \\
                                & 2,048 & 20.0  & 20.7 & 33.4M \\
           	\hline
           	\multirow{4}*{QRNN} & 256  & 23.6 & 23.9  &  0.6M \\
                                & 512 & 19.2 & 20.1  &  1.4M \\
           	                    & \textbf{1,024} & \textbf{17.4} & \textbf{18.5}  &  \textbf{3.8M} \\
                                & 2,048 & 17.5 & 18.7  & 11.2M \\
           	\hline
        \end{tabular}
        \label{table:timit2}
        }
\end{table}

The reported results show that the QRNN is a better framework for ASR systems than real-valued RNN when dealing with conventional multidimensional acoustic features. Indeed, the QRNN performed better and with less parameters, leading to a more efficient representation of the information.

%
%
\section{Conclusion}

\textbf{Summary.}
This paper proposes to investigate novel quaternion-valued architectures in two different conditions of speech recognition on the TIMIT phoneme recognition tasks. The experiments show that quaternion approaches always outperform real-valued equivalents in both benchmarks, with a maximum reduction factor of the number of learning parameters of $3.96$ times. It has been shown that the appropriate multidimensional quaternion representation of acoustic features, alongside with the \textit{Hamilton product}, help QCNN and QRNN to well-learn both internal and external relations that exists within the features, leading to a better generalization capability, and to a more efficient representation of the relevant information through significantly less free parameters than traditional real-valued neural networks. 

\textbf{Future Work.}
Future investigation will be to develop multi-view features that contribute to decrease ambiguities in representing phonemes in the quaternion space. In this extend, a recent approach based on a quaternion Fourrier transform to create quaternion-valued signal has to be investigated.

%
%
 
\bibliographystyle{plain}

\end{document}